\documentclass[preprint,showpacs,preprintnumbers,amsmath,amssymb]{revtex4-1}
\usepackage{latexsym}
\usepackage{epsfig}
\usepackage{float}
\usepackage{lipsum}
\usepackage{graphicx,times}
\usepackage{amssymb,amsmath}
\usepackage{color}
\usepackage{dcolumn}
\usepackage{epstopdf}
\usepackage{pgfplots}
\usepackage[colorlinks=true, urlcolor=blue, linkcolor=blue, citecolor=blue]{hyperref}
\newcommand{\be}{\begin{equation}}
\newcommand{\ee}{\end{equation}}

\begin{document}
\title{Signature of $N^{\star}$ resonance in mass spectrum of the $K\bar{K}N$ decay channels}
\author{Sajjad Marri}
\affiliation{Department of  Physics, Isfahan University of Technology, Isfahan 84156-83111, Iran}
\date{\today}
\begin{abstract}
Three-body calculations of $I=\frac{1}{2}$, $J^{\pi}=\frac{1}{2}^{+}$ state of 
$K\bar{K}N$ system were performed. Using separable potentials for two-body 
interactions in the Faddeev equation, different reaction process for $K\bar{K}N$ 
three-body system were studied. Within this method, the $\pi\Sigma{K}$ and 
$\pi\eta{N}$ mass spectra were extracted. Different types of $\bar{K}N-\pi\Sigma$ 
potentials based on phenomenological and chiral SU(3) approach were used and 
the dependence of the mass spectra to the $\bar{K}N$ model of interaction were 
studied. It was shown that the $\pi\Sigma{K}$ and $\pi\eta{N}$ mass spectra could 
be a useful tool to study the properties of the $N^{\star}$ resonance. 
\end{abstract}
\maketitle
\section{INTRODUCTION}
\label{intro}

The study of few-body systems, including antikaon is an important issue in contemporary 
strangeness nuclear physics and has attracted continuous attention. The $\bar{K}N$ system 
is a building-block of $\bar{K}$ nuclear few-body systems. This is based on the fact that 
the $\bar{K}N$ interaction is strongly attractive and the binding energy of the $\bar{K}N$ 
quasi-bound state is about $10\sim 30$ MeV~\cite{a1,a2,a3,a4,a5,a6,a7,a8,a9,a10,a11}. Of 
course, the binding energy of the antikaonic systems is not so large in comparison with 
the typical hadron energy scale and also the distance between constituent hadrons is larger than 
the typical size of them. Consequently, all hadronic constituents will keep their identity. 
Such quasi-bound states are usually called as hadronic molecular states~\cite{h1}. During 
the last two decades, various hadronic molecular states, including $\bar{K}$ and $K$ mesons 
were studied by different groups~\cite{d1,d2,d3,d4,d5,d6,d7}. Among these molecular states, 
the $\bar{K}NN$ with quantum numbers $I=\frac{1}{2}$ and $J^{\pi}=0^{-}$ bound state has been 
more the object of intense theoretical~\cite{b1,b2,b3,b4,b5,b6,b7,b8,b9,b10,b11,b12,b13,b14} 
and experimental~\cite{c1,c2,c3,c4,c5,c6,c7} studies.

Plus the $\bar{K}NN$, numerous theoretical works were also performed to study systems 
of two mesons and one baryon with strangeness $S=0$, finding resonant states which could 
be identified with the existing baryonic resonances~\cite{d1,d2,d3,d4}. A variational 
calculation of the three-body $K\bar{K}N$ system was performed in Ref.~\cite{d1} using 
effective potentials for $\bar{K}N$ and $K\bar{K}$ interactions. The $\Lambda(1405)$ 
resonance is generated as quasi-bound state in $\bar{K}N$ system and the scalar mesons 
$f_{0}(980)$ and $a_{0}(980)$ are reproduced as quasi-bound states of $K\bar{K}$ system 
in $I=0$ and $I=1$ isospin channels, respectively. In this calculation, a quasi-bound 
state ($N^{\star}$ resonance) with quantum numbers $I=\frac{1}{2}$ and $J^{\pi}=\frac{1}{2}^{+}$ 
was found with a mass 1910 MeV and a width 90 MeV below all of the meson-baryon decay 
threshold energies of the $\Lambda(1405)+K$, $f_{0}(980)+N$ and $a_{0}(980)+N$ states. 
It was concluded that the $K\bar{K}N$ state can be understood by the structure of 
simultaneous coexistence of $\Lambda(1405)+K$ and $a_{0}(980)+N$ clusters and the 
$\bar{K}$ meson is shared by both $\Lambda(1405)$ and $a_{0}(980)$ at the same time. 
This quasi-bound state was also studied in Refs.~\cite{d2,d3} by using coupled-channel 
relativistic Faddeev equation and in Ref.~\cite{d4} using fixed center approximation 
of three-body Faddeev equation. The extracted pole energies were in agreement with the 
pole in Ref.~\cite{d1}. A discussion on experimental observation of this $N^{\star}$ 
can be also found in Ref.~\cite{d8}.

\begin{figure}[H]
\centering
\includegraphics[width=8.6cm]{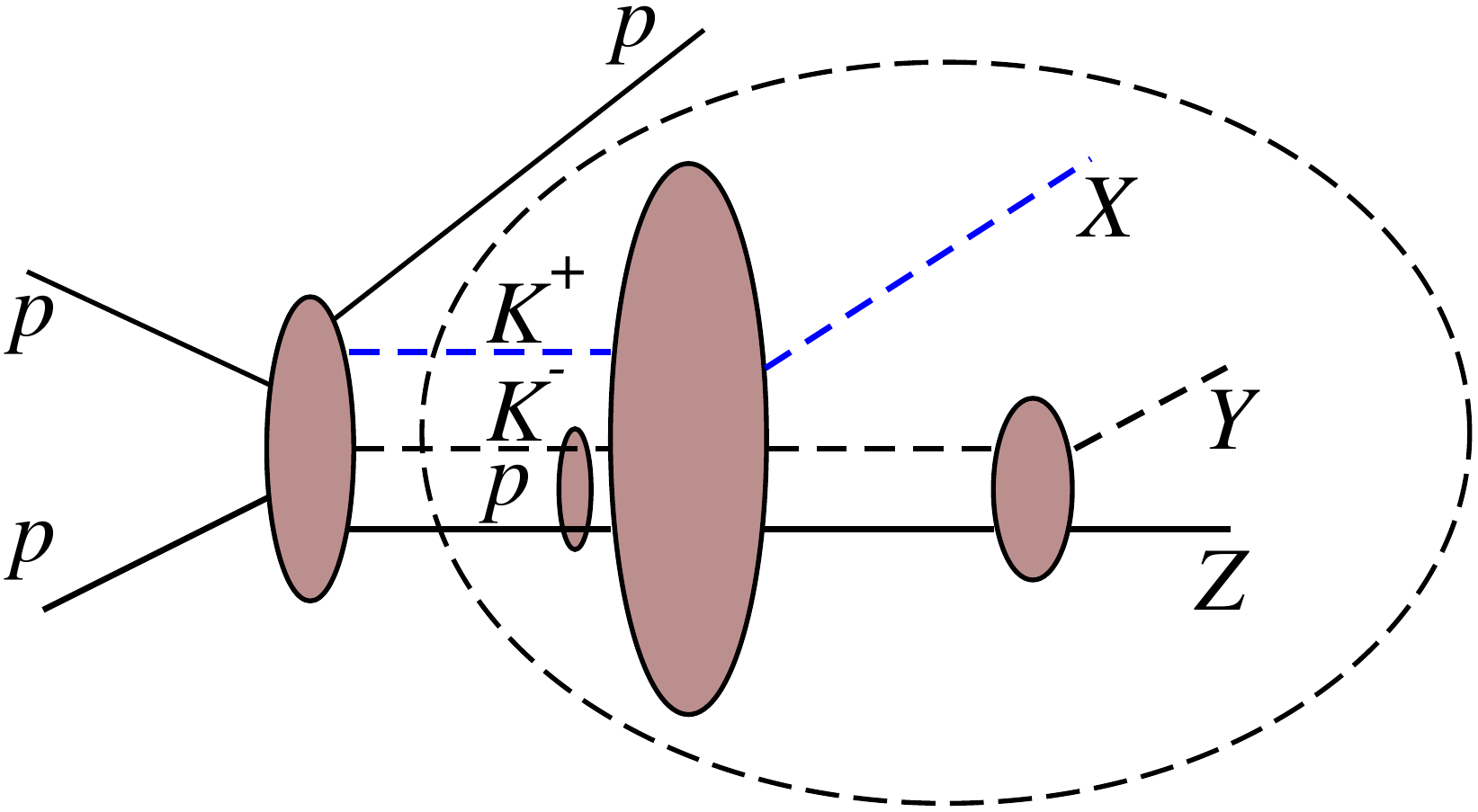} \\
\caption{(Color online) Diagram for proton-proton reaction and 
formation of the $K\bar{K}N$ system. In decay channel, when $X$ 
is equal to $K^{+}$ the $YZ$ pair is $\pi\Sigma$ and when $X$ 
is equal to $p$ the $YZ$ pair should be $\pi\pi$ and $\pi\eta$ 
depending on the total isospin of the $YZ$ pair.}
\label{fig.1}
\end{figure}

The $K\bar{K}N$ quasi-bound state can be produced in $\gamma{p}$ and $pp$ reactions, 
and the signal of the resonance may be observed in the mass spectrum of the final 
particles~\cite{d8}. The investigation for the $K\bar{K}N$ quasi-bound state was explored 
through $\gamma$ incident reaction $\gamma{p}\rightarrow K^{+}\Lambda$ 
by CLAS experiment at JLab~\cite{h2,h3}. The $K\bar{K}N$ quasi-bound state could be 
also studied through the $pp$ reaction (see Fig.~\ref{fig.1}). This reaction was 
performed as a HADES experiment at GSI~\cite{h4}. 
This paper is devoted to investigating the pole structure of the $K\bar{K}N$ three-body 
system. It was studied how well the signal of the $K\bar{K}N$ quasi-bound state can be 
observed in the $\pi\Sigma{K}$ and $\pi\eta{N}$ mass spectrum resulting from reaction 
under consideration. The few-body calculations for the $K\bar{K}N$ system were performed 
by using Faddeev AGS equations~\cite{f1}. The transition probabilities for the 
$(\bar{K}N)_{I=0}+K$ and $(K\bar{K})_{I=1}+N$ reactions were calculated. Within this 
method, the behavior of the transition probability was investigated. Different 
phenomenological and chiral based $\bar{K}N-\pi\Sigma$ potentials were used~\cite{e2,e3} 
to investigate the sensitivity of the three-body observables on two-body inputs.

The paper is organized as follows: in Sect.~\ref{formula}, I will explain the formalism 
used for the three-body $K\bar{K}N$ system and give a brief description of the transition 
probability formula for reaction under consideration. Sect.~\ref{results} is devoted to 
the two-body inputs of the calculations and representation of the computed pole energies 
and mass spectra. In Section~\ref{conc}, I give conclusions.
\section{FORMALISM AND BASIC INGREDIENTS}
\label{formula}
The present three-body calculations of the $K\bar{K}N$ system are based on the AGS 
form of the Faddeev equation~\cite{f1}. To describe the two-body interactions, 
which are the basic ingredient of the calculations, separable potentials with 
the following form were used
\begin{equation}
V^{I}_{\alpha\beta}(k_{\alpha},k_{\beta})=g^{I}_{\alpha}(k_{\alpha})\, 
\lambda^{I}_{\alpha\beta} \, g^{I}_{\beta}(k_{\beta}),
\label{eq.1}
\end{equation}
where the quantities $g^{I}_{\alpha}(k_{\alpha})$ are the form factors of the interacting 
two-body subsystem with relative momentum $k_{\alpha}$ and isospin $I$. The strength 
parameters of the interaction are denoted by $\lambda^{I}_{\alpha\beta}$. To include the 
low-lying channels in $\bar{K}N$ and $K\bar{K}$ interactions, the potentials are labeled 
also by $\alpha$ and $\beta$ indexes. The separable form of the two-body $T$-matrices is 
given by
\begin{equation}
T^{I}_{\alpha\beta}(k_{\alpha},k_{\beta};z)=g^{I}_{\alpha}(k_{\alpha})\, 
\tau^{I}_{\alpha\beta}(z)\, g^{I}_{\beta}(k_{\beta}),
\label{eq.2}
\end{equation}
where the operator $\tau^{I}_{\alpha\beta}(z)$ is the usual two-body propagator and $z$ 
is the two-body energy. There are three different particles in the system under consideration. 
Therefore, defining the interacting pairs and their allowed spin and isospin quantum 
numbers, the $K\bar{K}N$ three-body system will have the following partitions
\begin{equation}
\begin{split}
& (1):(\bar{K}N)_{s=\frac{1}{2};I=0,1}+K, \\ 
& (2):(KN)_{s=\frac{1}{2};I=0,1}+\bar{K}, \\
& (3):(K\bar{K})_{s=0;I=0,1}+N.
\end{split}
\label{eq.3}
\end{equation}

\begin{figure}[H]
\centering
\includegraphics[width=8.6cm]{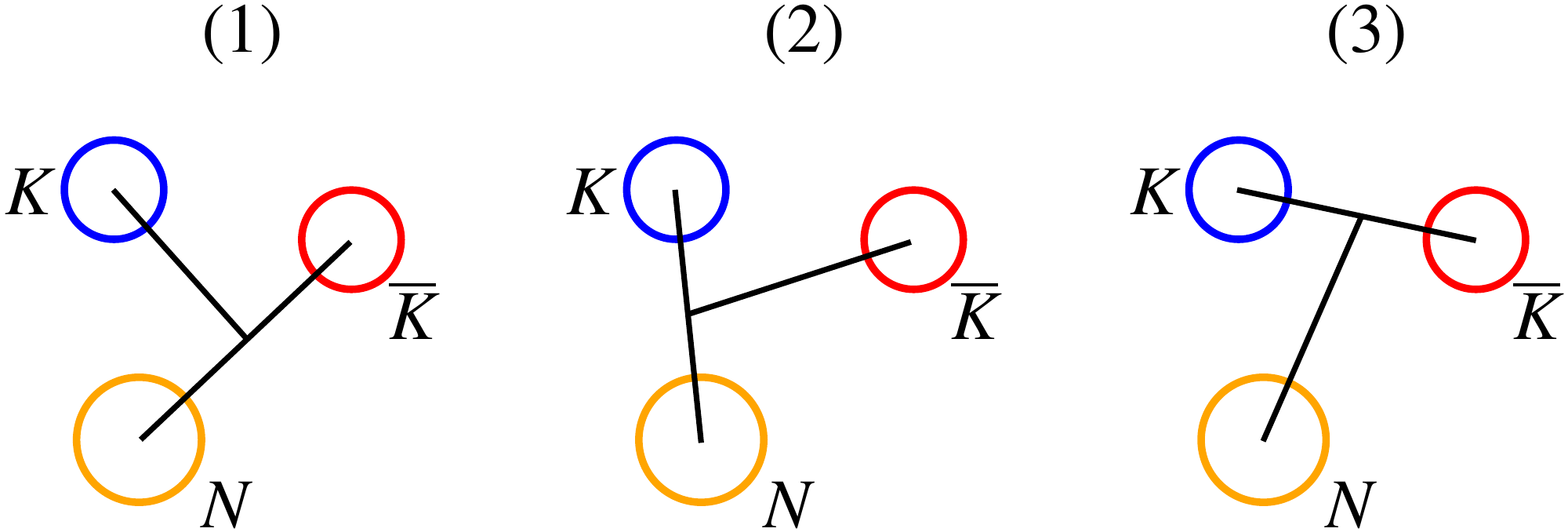} \\
\caption{(Color online) Diagrammatic representation of different 
partitions of the $K\bar{K}N$ system. Defining the interacting 
particles, there are three partitions, namely $K+(\bar{K}N)$, 
$\bar{K}+(KN)$ and $(K\bar{K})+N$. The antikaon is defined by 
red circle, the kaon by blue circle and the nucleon by orange circle.}
\label{fig.2}
\end{figure}

For convenience, one can introduce the three rearrangement channels $i=1,2,3$ of the 
$K\bar{K}N$ three-body system as shown in Fig.~\ref{fig.2}. The quantum numbers of the 
$K\bar{K}N$ are $I=\frac{1}{2}$ and $s=\frac{1}{2}$. Therefore, in actual calculations, 
when one includes isospin and spin indexes the number of configurations is equal to six, 
corresponding to different possible two-quasi-particle partitions. Using separable 
potential for two-body interactions, the three-body Faddeev equation~\cite{b5} in the 
AGS form is given by
\begin{equation}
\begin{split}
& \mathcal{K}_{i,j}^{I_{i},I_{j}}(p_{i},p_{j};W)= 
\mathcal{M}_{i,j}^{I_{i},I_{j}}(p_{i},p_{j};W)
+\sum_{k,I_{k}}\int {d}^{3}p_{k} \\
& \hspace{0.5cm}\times\mathcal{M}_{i,k}^{I_i,I_k}(p_{i},p_{k};W) 
\, \tau_{k}^{I_k}(W-\frac{p^{2}_{k}}{2\nu_{k}})
\, \mathcal{K}_{k,j}^{I_k,I_j}(p_{k},p_{j};W).
\end{split}
\label{eq.4}
\end{equation}

Here, the operators $\mathcal{K}_{ij}^{I_{i}I_{j}}$ are the transition amplitudes 
which describe the elastic and re-arrangement processes 
$i+(jk)_{I_{i}}\rightarrow j+(ki)_{I_{j}}$~\cite{b5} and the operators 
$\mathcal{M}_{ij}^{I_{i}I_{j}}$ are the corresponding Born terms. In this equation 
$W$ is the three-body energy and $W-\frac{p^{2}_{k}}{2\nu_{k}}$ is the energy of 
the interacting pair $(ij)$ where, $\nu_{i}=m_{i}(m_{j}+m_{k})/(m_{i}+m_{j}+m_{k})$, 
is the reduced mass, when particle $i$ is a spectator. Faddeev partition indexes 
$i,j,k=1,2,3$ denote simultaneously an interacting pair and a spectator particle. 
Depending on the spectator particle, the operator 
$\tau_{k}^{I_{k}}(W-\frac{p^{2}_{k}}{2\nu_{k}})$, $k=1,2,3$ is given by
\begin{equation}
\begin{split}
& \tau_{1}^{I_{1}}=\tau_{K}^{I_{K}}=
\tau_{\bar{K}N-\bar{K}N}^{I_{\bar{K}N}}(W-\frac{p^{2}_{K}}{2\nu_{K}}), \\
& \tau_{2}^{I_{2}}=\tau_{\bar{K}}^{I_{\bar{K}}}=
\tau_{KN-KN}^{I_{KN}}(W-\frac{p^{2}_{\bar{K}}}{2\nu_{\bar{K}}}), \\
& \tau_{3}^{I_{3}}=\tau_{N}^{I_{N}}=
\tau_{K\bar{K}-K\bar{K}}^{I_{K\bar{K}}}(W-\frac{p^{2}_{N}}{2\nu_{N}}).
\end{split}
\label{eq.44}
\end{equation}

The $\bar{K}N$ system is coupled with $\pi\Sigma$ channel and the $K\bar{K}$ system 
is coupled with $\pi\pi$ and $\pi\eta$ systems in isospin $I=0$ and $I=1$ channels, 
respectively. In actual calculations, one should extend the Faddeev equation to include 
the low-lying channels. In present calculations, the $\bar{K}N-\pi\Sigma$ and 
$K\bar{K}-\pi\pi-\pi\eta$ couplings are not taken into account directly and the particle 
indexes are omitted for all three-body operators and the interactions between particles 
in low-lying channels are neglected. Thus, our coupled-channels three-body calculations 
with coupled-channel $\bar{K}N-\pi\Sigma$ and $K\bar{K}-\pi\pi-\pi\eta$ interactions are 
equivalent to the one-channel three-body calculation using the so-called 
\textquotedblleft{exact optical}\textquotedblright $\bar{K}N(-\pi\Sigma)$ and 
$K\bar{K}(-\pi\pi-\pi\eta)$ potentials~\cite{e1}. The decay to the $K\pi\Sigma$, 
$\pi\eta{N}$ and $\pi\pi{N}$ channels is taken into account through the imaginary part 
of the optical $\bar{K}N(-\pi\Sigma)$ and $K\bar{K}(-\pi\pi-\pi\eta)$ potentials. 

Supposing the $(\bar{K}N)_{I=0}+K$ as the initial state of the $K\bar{K}N$ system, the 
three-body Faddeev AGS equations can be given by 
\begin{equation}
\begin{split}
& \mathcal{K}_{K,K}^{I_{{K}},0} =  \mathcal{M}_{K,N}^{I_{K},I_{N}}
\tau_{N}^{I_{N}}\mathcal{K}_{N,K}^{I_{N},0}
 + \mathcal{M}_{K,\bar{K}}^{I_{K},I_{\bar{K}}}
\tau_{\bar{K}}^{I_{\bar{K}}} \mathcal{K}_{\bar{K},K}^{I_{\bar{K}},0}, \\
& \mathcal{K}_{N,K}^{I_{N},0} = \mathcal{M}_{N,K}^{I_{N},0} + 
\mathcal{M}_{N,K}^{I_{N},I_{K}}\tau_{K}^{I_{K}}
\mathcal{K}_{K,K}^{I_{K},0} + 
\mathcal{M}_{N,\bar{K}}^{I_{N},I_{\bar{K}}}\tau_{\bar{K}}^{I_{\bar{K}}}
\mathcal{K}_{\bar{K},K}^{I_{\bar{K}},0}, \\
& \mathcal{K}_{\bar{K},K}^{I_{\bar{K}},0}  =  
\mathcal{M}_{\bar{K},K}^{I_{\bar{K}},0} + 
\mathcal{M}_{\bar{K},K}^{I_{\bar{K}},I_{K}}\tau_{K}^{I_{K}}\mathcal{K}_{K,K}^{I_{K},0} + 
\mathcal{M}_{\bar{K},N}^{I_{\bar{K}},I_{N}}\tau_{N}^{I_{N}}\mathcal{K}_{N,K}^{I_{N},0}. \\
\end{split}
\label{eq.5}
\end{equation}

To study the possible signature of the $N^{\star}$ resonance in the mass spectrum of the 
final particles in $K+(\bar{K}N)_{I=0}$ reaction, three different channels can be studied 
which are given by 
\begin{equation}
\begin{split}
& K+(\bar{K}N)_{I=0}\rightarrow{K}\pi\Sigma,\\
& K+(\bar{K}N)_{I=0}\rightarrow\pi\eta{N},\\
& K+(\bar{K}N)_{I=0}\rightarrow\pi\pi{N},
\end{split}
\label{eq.66}
\end{equation}
where the first two reactions are more probable~\cite{d1}. To define the $K\pi\Sigma$ and 
$\pi\eta{N}$ mass spectrum, first one should define break-up amplitude. Since, the low-lying 
channels are not directly included in the calculations, the only Faddeev amplitudes which 
contribute in the scattering amplitude are $\mathcal{K}_{K,K}^{I,0}(p_{K},\bar{P}_{K};W)$ 
and $\mathcal{K}_{N,K}^{1,0}(p_{N},\bar{P}_{K};W)$ for extracting the $K\pi\Sigma$ and 
$\pi\eta{N}$ mass spectrum, respectively. Therefore, the scattering amplitude can be expressed as 
\begin{equation}
\begin{split}
& T_{(\pi\Sigma)+{K}\leftarrow(\bar{K}N)_{I=0}+K} (\vec{k}_{K},\vec{p}_{K},\bar{P}_{K};W)
=\sum_{I}g_{\pi\Sigma}^{I}(\vec{k}_{K} )\, \\
& \hspace{1.5cm}\times \tau_{K}^{I}(W-\frac{p^{2}_{K}}{2\nu_{K}}) 
 \mathcal{K}_{K,K}^{I,0}(p_{K},\bar{P}_{K};W),\\
& T_{(\pi\eta)+{N}\leftarrow(\bar{K}N)_{I=0}+K} (\vec{k}_{N},\vec{p}_{N},\bar{P}_{K};W)
=g_{\pi\eta}^{I=1}(\vec{k}_{N} )\, \\
& \hspace{1.5cm}\times \tau_{N}^{I=1}(W-\frac{p^{2}_{N}}{2\nu_{N}})
\mathcal{K}_{N,K}^{1,0}(p_{N},\bar{P}_{K};W),
\end{split}
\label{eq.6}
\end{equation}
where, $\vec{k}_{i}$ is the relative momentum between the interacting pair ($jk$) and 
$\bar{P}_{K}$ is the initial momentum of spectator $K$ in $K\bar{K}N$ center of mass. 
The quantities $\mathcal{K}_{i,j}^{I_{i},I_{j}}$ are the Faddeev amplitudes, which are 
derived from Faddeev equation (\ref{eq.5}). 

Using Eq.(\ref{eq.6}), the transition probability of $K+(\bar{K}N)_{I=0}$ reaction can 
be defined as follows,
\begin{equation}
\begin{split}
& w_{1}(\bar{P}_{K},W)=\int d^{3}p_{K}\int d^{3}k_{K}\, \delta(W-Q_{1}(p_{K},k_{K})) \\
& \hspace{1.5cm}\times|T_{(\pi\Sigma)+{K}\leftarrow(\bar{K}N)_{I=0}+K}(\vec{k}_{K},
\vec{p}_{K},\bar{P}_{K};W)|^2, \\
& w_{2}(\bar{P}_{K},W)=\int d^{3}p_{N}\int d^{3}k_{N}\, \delta(W-Q_{2}(p_{N},k_{N})) \\
& \hspace{1.5cm}\times|T_{(\pi\eta)+{N}\leftarrow(\bar{K}N)_{I=0}+K}(\vec{k}_{N},
\vec{p}_{N},\bar{P}_{K};W)|^2,
\end{split}
\label{eq.7}
\end{equation}
where $Q_{1}(p_{K},k_{K})$ and $Q_{2}(p_{N},k_{N})$ are given by
\begin{equation}
\begin{split}
& Q_{1}(p_{K},k_{K})=\frac{p^{2}_{K}(m_{K}+m_{\pi}+m_{\Sigma})}{2m_{K}(m_{\pi}+m_{\Sigma})}-
\frac{k^{2}_{K}(m_{\pi}+m_{\Sigma})}{2m_{\pi}m_{\Sigma}}, \\
& Q_{2}(p_{N},k_{N})=\frac{p^{2}_{N}(m_{N}+m_{\pi}+m_{\eta})}{2m_{N}(m_{\pi}+m_{\eta})}-
\frac{k^{2}_{N}(m_{\pi}+m_{\eta})}{2m_{\pi}m_{\eta}}.
\end{split}
\label{eq.8}
\end{equation}
\section{Results and discussions}
\label{results}
Before I proceed to present the results, I should give a brief description of the 
two-body interactions, which are used in present calculations. The $K\bar{K}N$ 
three-body system have three different subsystems, which are $\bar{K}N$, $K\bar{K}$ 
and $KN$. The $\bar{K}N$ subsystem is coupled with $\pi\Sigma$ and $K\bar{K}$ subsystem 
is coupled with $\pi\pi$ and $\pi\eta$ in $I=0$ and $I=1$ channels, respectively. 
Therefore, in full calculation of $K\bar{K}N$ system, the $K\pi\Sigma$, $\pi\pi{N}$ 
and $\pi\eta{N}$ channels should be included too. In the present work, the low-lying 
channels are not included directly. Therefore, the $\pi{N}$, $\pi{K}$, $\eta{N}$ and 
$\Sigma{K}$ interactions in low-lying channels are neglected and the decay to 
$K\pi\Sigma$, $\pi\pi{N}$ and $\pi\eta{N}$ is included by using the so-called exact 
optical potentials. In the following, I will give a brief description of the 
$\bar{K}N-\pi\Sigma$, $K\bar{K}-\pi\pi-\pi\eta$ and $KN$ interactions.

The $\bar{K}N$ interaction which is the fundamental ingredient to study few-body systems 
with antikaon is closely related to the structure of $\Lambda(1405)$ resonance in the isospin 
zero channel. The $\Lambda(1405)$ is located slightly below the $\bar{K}N$ threshold and 
decays into the $\pi\Sigma$ channel through the strong interaction. The $\Lambda(1405)$ 
resonance can be interpreted as a quasi-bound state of $\bar{K}N$ with a binding energy 
of $27$ MeV~\cite{a6}. On the other hand, the theoretical calculation based on chiral SU(3) 
dynamics claim that $\Lambda(1405)$ is dynamically generated by the meson-baryon 
interactions and consists of two poles coupled to the $\pi\Sigma$ and $\bar{K}N$ states~\cite{a3,a8}. 
According to the chiral models, the pole position in the $S$-wave $\bar{K}N$ scattering 
amplitude is located at $\sim{1426}$ MeV. Thus, the $\bar{K}N$ interaction is expected to 
be weaker than that predicted by other phenomenological model calculations. In present 
calculation, different models of interaction were used to describe the coupled channel 
$\bar{K}N-\pi\Sigma$ system. Two different phenomenological plus one energy-dependent 
chiral based potential for $\bar{K}N$ interaction~\cite{e2,e3} were used. The potentials 
yield the one- and two-pole structure of the $\Lambda(1405)$ resonance. The parameters of the 
phenomenological potentials, are given in Ref.~\cite{e2}. From now, I refer these potentials 
as $\mathrm{SIDD}^{1}$ and $\mathrm{SIDD}^{2}$ which, have the one- and two-pole structure of 
the $\Lambda(1405)$ resonance, respectively. The SIDD potentials are adjusted to reproduce the 
results of the SIDDHARTA experiment. The parameters of the energy-dependent chiral potential 
are given in Ref.~\cite{e3}.
\begin{table*}[t!]
\caption{Range parameters $\Lambda_{K\bar{K}}^{I}$ (in $\mathrm{fm}^{-1}$) 
and strength parameters $\lambda_{\alpha\beta}^{I}$ (in $\mathrm{fm}^{-2}$) 
of the $K\bar{K}-\pi\pi-\pi\eta$ potential. The range parameters in each 
isospin channel are independent of two-body channel. For the $\alpha$ and 
$\beta=K\bar{K}$, we have $\alpha=\beta=1$ and for $\alpha=\beta=\pi\pi$ 
and $\alpha=\beta=\pi\eta$, the value of the $\alpha$ and $\beta$ is equal to two.}
\centering
\begin{tabular}{ccccccccc}
\hline\hline\noalign{\smallskip}\noalign{\smallskip}
&\,  $\Lambda_{K\bar{K}}^{0}$ \,&\,  $\Lambda_{K\bar{K}}^{1}$ \, 
& \, $\lambda_{11}^{0}$ \, & \, $\lambda_{12}^{0}$ \, & \, $\lambda_{22}^{0}$ \,
& \, $\lambda_{11}^{1}$ \, & \, $\lambda_{12}^{1}$ \, & \, $\lambda_{22}^{1}$  \\
\noalign{\smallskip}\noalign{\smallskip}\hline\noalign{\smallskip}\noalign{\smallskip}
$V^{I}_{K\bar{K}}$ \, & \, $3.570$ \, & \, $3.396$ \, 
& \, $-1.757$ \, & \, $2.975$ \, & \, $1.187$ \,
& \, $-1.503$ \, & \, $1.922$ \, & \, $0.092$ \\
\noalign{\smallskip}\noalign{\smallskip}\hline\hline
\end{tabular}
\label{ta.1} 
\end{table*}

The $K\bar{K}$ system is coupled with $\pi\pi$ and $\pi\eta$ in $I=0$ and $I=1$ channels, 
respectively. A coupled-channel potential was constructed for $K\bar{K}$ interaction in 
both isospin channels to take into account the decay of the $K\bar{K}$ system to $\pi\pi$ 
and $\pi\eta$ channels. A separable potential in the form given in Eq.~\ref{eq.1} were used 
to describe the $K\bar{K}-\pi\pi-\pi\eta$ interaction. To define the strength parameter 
$\lambda_{\alpha\beta}^{I}$, I used the pole energy of $f_{0}(980)$ and $a_{0}(980)$ 
resonances and also the $K\bar{K}$ scattering length~\cite{e4}. To determine the parameters 
of the $K\bar{K}$ interaction in both $I=0$ and $I=1$ channels, the mass 980 MeV and the 
width 60 MeV were taken for $f_{0}$ and $a_{0}$ resonances, which are close to the reported 
mass and width by PDG~\cite{e5}. The extracted parameters of the $K\bar{K}$ potential in the 
$I=0$ and $I=1$ channels, are given in Table~\ref{ta.1} and the form factors are taken to be 
in Yamaguchi form~\cite{y1}.

To describe the repulsive $KN$ interaction, I used a one-channel real potential in the 
form
\begin{equation}
\begin{split}
& V^{I}_{KN}(k,k')=g^{I}_{KN}(k)\, \lambda^{I}_{KN}\, g^{I}_{KN}(k'), \\
& g^{I}_{KN}(k)=\frac{1}{\Lambda_{KN}^{2}+k^{2}}.
\end{split}
\label{eq.9}
\end{equation}

The range parameter of the $KN$ potential, $\Lambda_{KN}$, was set to 3.9 $\mathrm{fm}^{-1}$. 
The $KN$ interaction with isospin $I=0$ is very weak. Therefore, it would not change the results 
of the present work and can be neglected. The strength parameter in $I=1$ channel is adjusted to 
reproduce the $KN$ scattering length. The experimental value of the scattering length for the 
$I=1$ channel is $a^{I=1}_{KN}=-0.310\pm 0.003\,\mathrm{fm}$~\cite{d1,e6,e7}. Therefore, the 
value of the strength parameter is $\lambda^{I=1}_{KN}=2.794\mathrm{fm}^{-2}$.
\subsection{Pole position of the three-body $K\bar{K}N$ system}
\begin{figure}[h!]
\centering
\includegraphics[width=8.6cm]{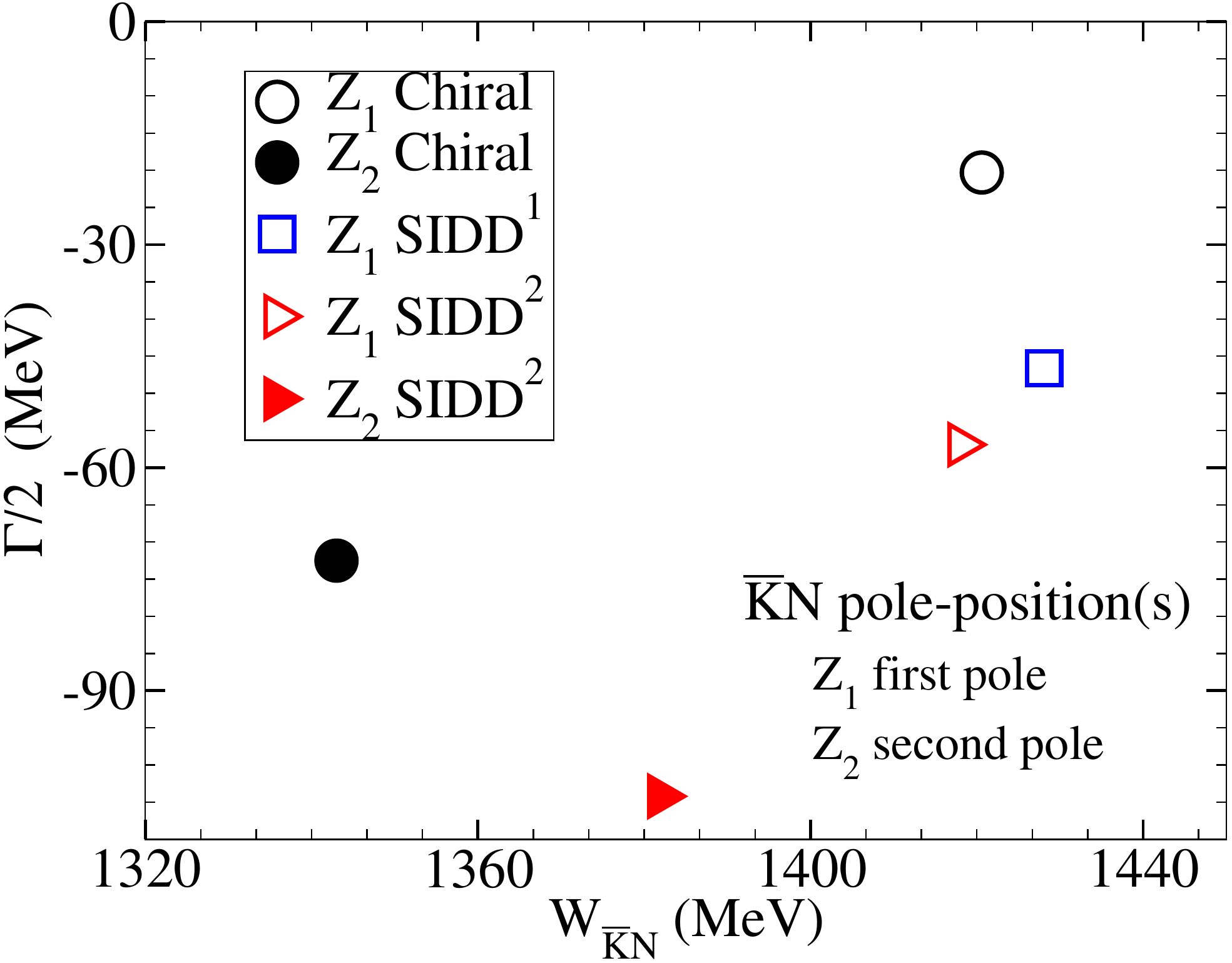} \\
\vspace{0.2cm}
\includegraphics[width=8.6cm]{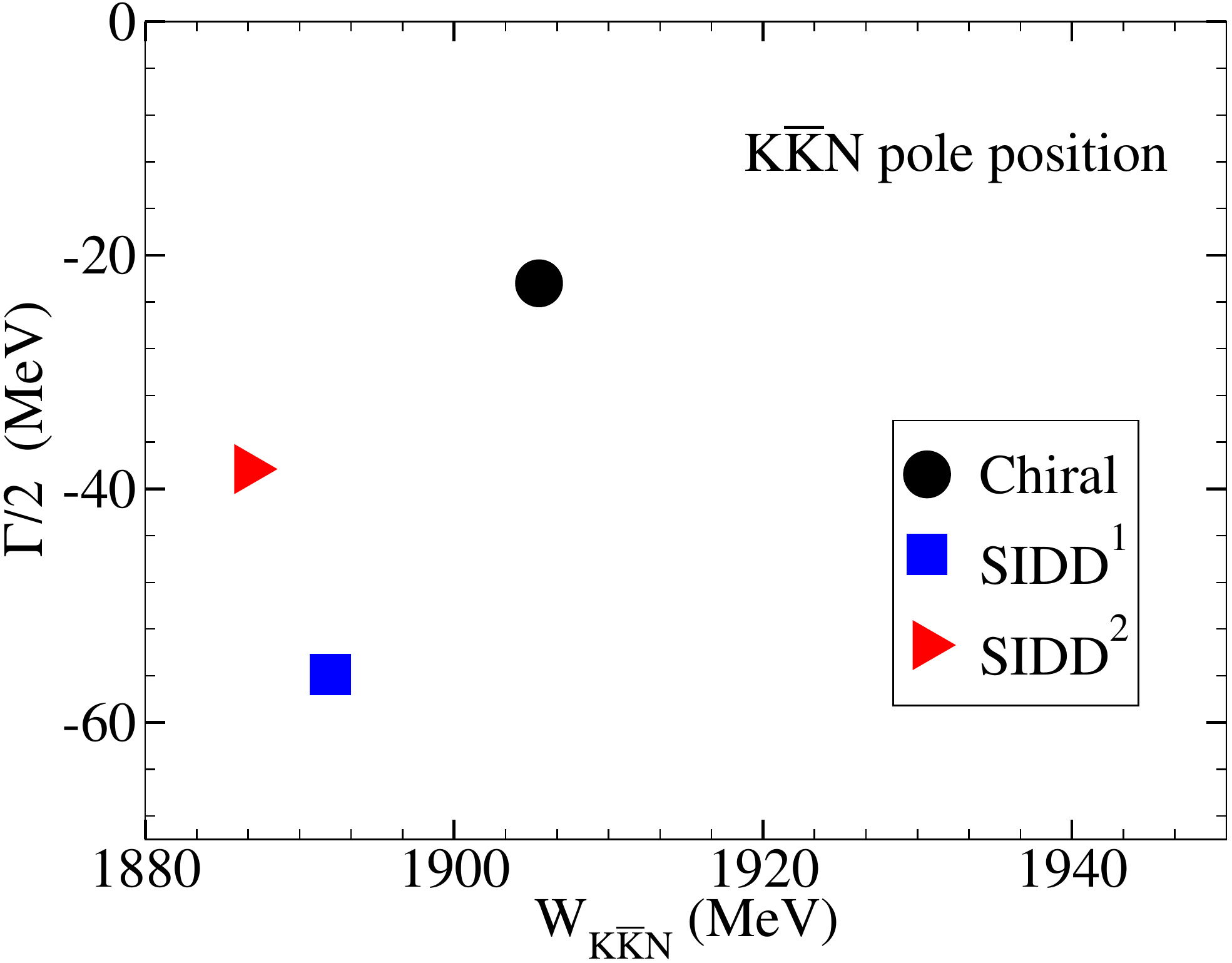} 
\caption{(Color online) The sensitivity of the pole position(s) 
(in MeV) of the $\bar{K}N$ (up) and $K\bar{K}N$ (down) systems 
to the different phenomenological and chiral based models of the 
$\bar{K}N-\pi\Sigma$ interaction is investigated. In the upper 
panel the quantities $Z_{1}$ and $Z_{2}$ standing for first and 
the second pole of $\Lambda$(1405) resonance, respectively.}
\label{fig.3}
\end{figure}

Before I proceed to represent the extracted mass spectra for $K+(\bar{K}N)_{I=0}$ 
reaction, in this subsection I shall begin with a survey on pole structure of the 
$K\bar{K}N$ system. The obtained results can be used as a guideline in interpreting 
the behavior of the extracted mass spectra from $K+(\bar{K}N)_{I=0}$ reaction. Solution 
of the Faddeev equation corresponding to bound states and resonance poles in the 
$(I,J^{\pi})=(\frac{1}{2},\frac{1}{2}^{+})$ channel of the $K\bar{K}N$ three-body 
system was found by solving the homogeneous version of the Faddeev AGS equation, 
which are defined by
\begin{equation}
\begin{split}
& u_{n;i}^{I_{i}}(p_{i};W)= \frac{1}{\lambda_{n}}\sum_{j,I_{j}}
\int {d}^{3}p_{j}(1-\delta_{ij})\,\mathcal{M}_{i,j}^{I_i,I_j}(p_{i},p_{j};W) \\
& \hspace{1.7cm}\times\tau_{j}^{I_j}(W-\frac{p^{2}_{j}}{2\nu_{j}})\, 
u_{n;j}^{I_{j}}(p_{j};W),
\label{eq.10}
\end{split}
\end{equation}
where $\lambda_{n}$ and the form factors $u_{n;i}^{I_i}(p_{i};W)$ are taken as 
the eigenvalues and eigenfunctions of the kernel of the equation (\ref{eq.4}), 
respectively. In Fig.~\ref{fig.3} (lower panel) the results of the present work 
for three-body $K\bar{K}N$ quasi-bound state are presented and sensitivity of 
the $K\bar{K}N$ pole position to the $\bar{K}N-\pi\Sigma$ interaction is 
investigated by using different potential models. Plus the $K\bar{K}N$ pole 
position, I also present the pole position of the quasi-bound states in the 
$\bar{K}N$ system for phenomenological and energy-dependent chiral potentials 
(upper panel).
\begin{figure*}
\centering
\includegraphics[width=8.cm]{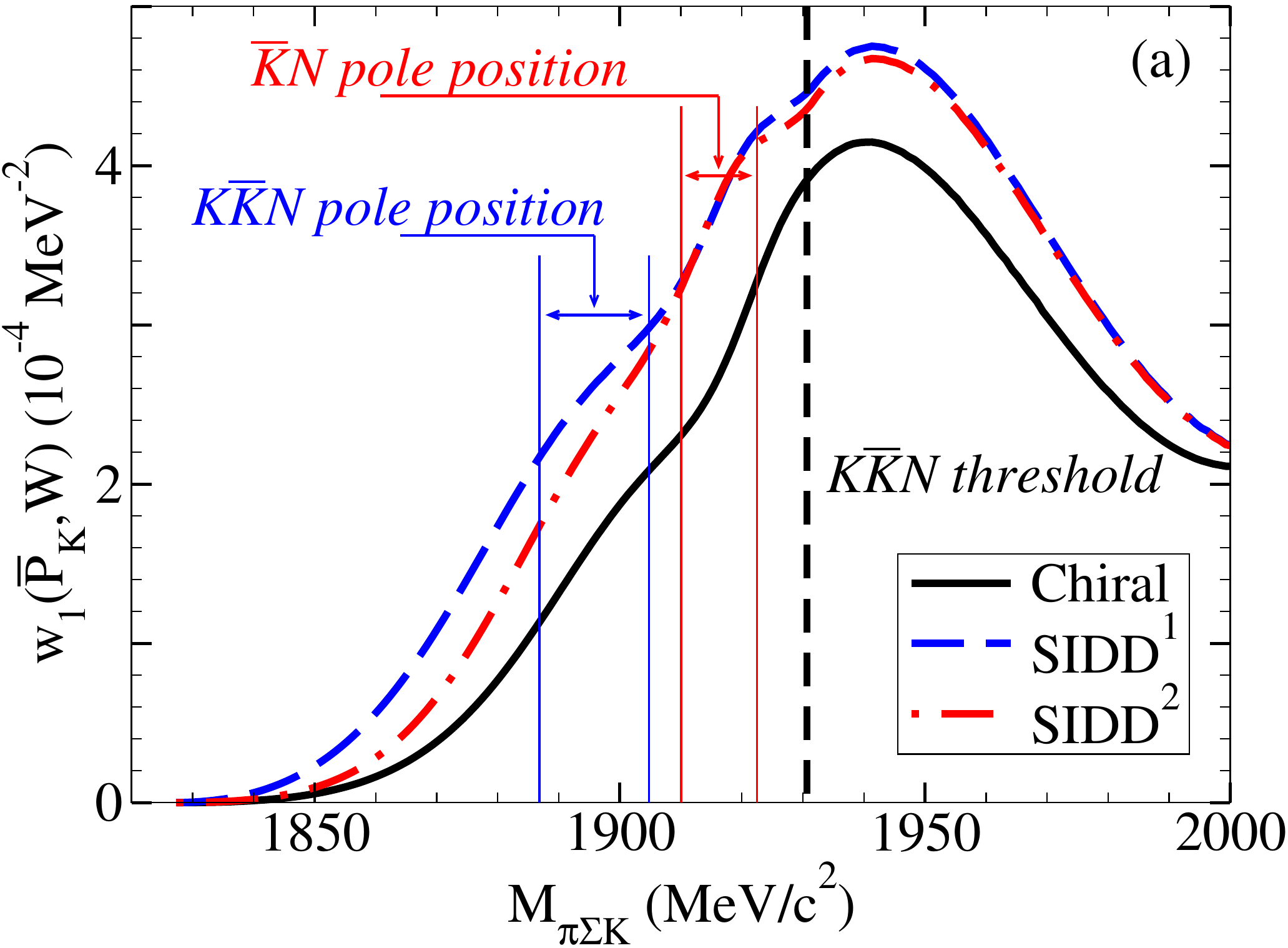}
\includegraphics[width=8.cm]{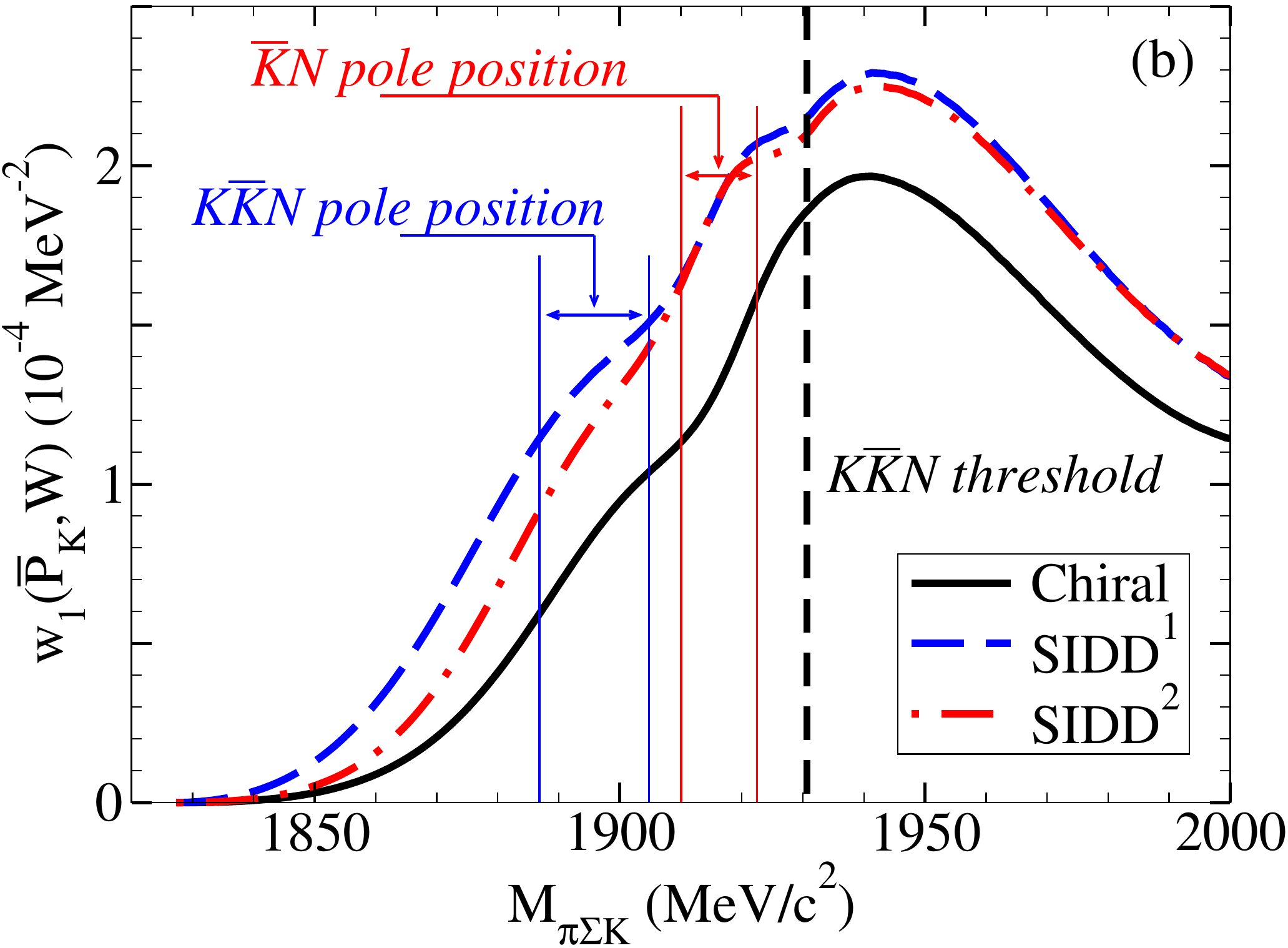} \\ 
\includegraphics[width=8.cm]{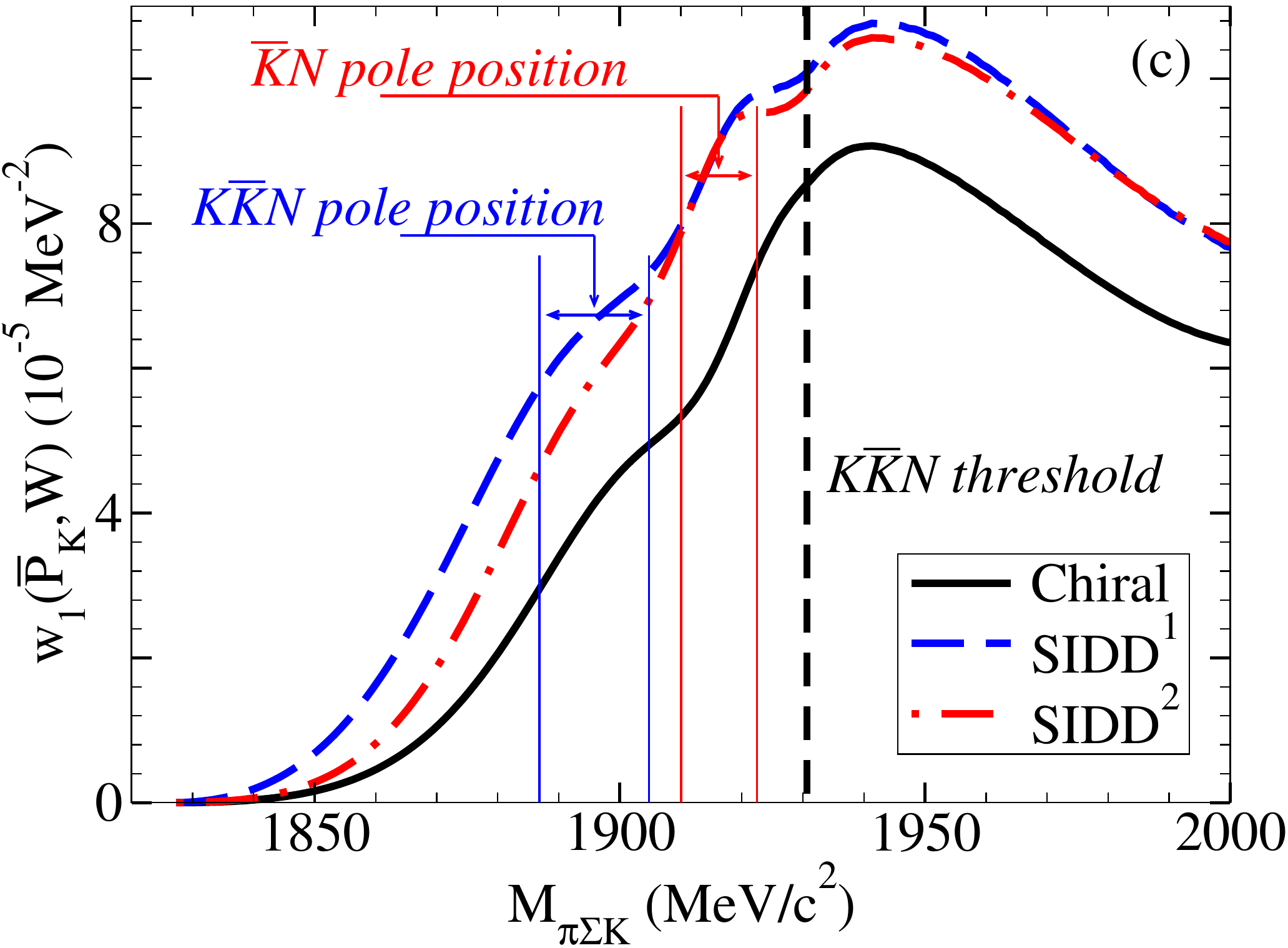}
\includegraphics[width=8.cm]{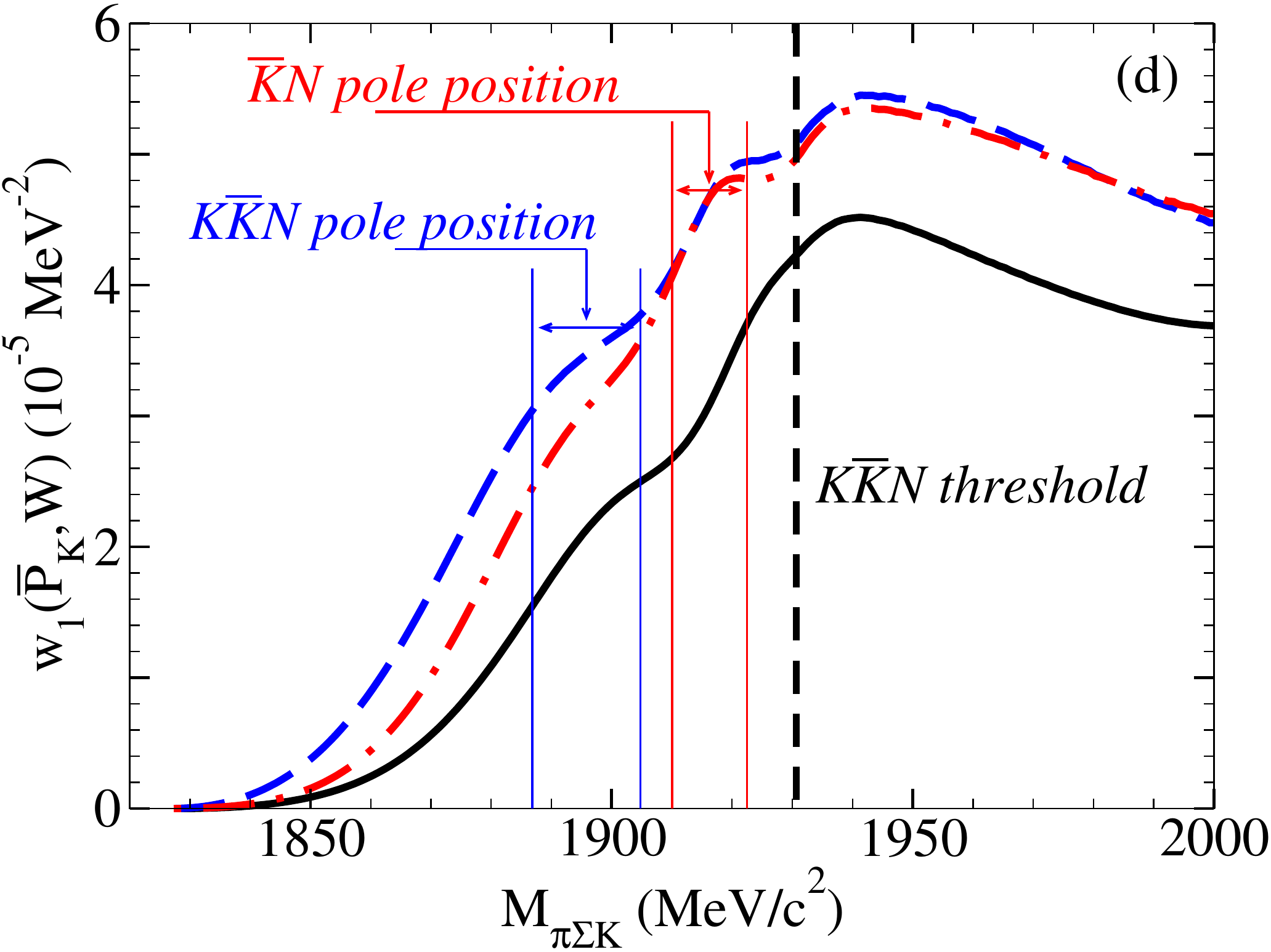}
\caption{(Color online) The $\pi\Sigma{K}$ mass spectra for 
$(\bar{K}N)_{I=0}+K\rightarrow\pi\Sigma{K}$ reaction. Different 
types of $\bar{K}N-\pi\Sigma$ potentials were used: the one- and 
two-pole version of the SIDD potentials~\cite{e2} and also one 
energy-dependent chiral potential~\cite{e3}. The transition 
probabilities are calculated for different values of $\bar{P}_{K}$. 
In panels (a), (b), (c) and (d), the values of $\bar{P}_{K}$ are 
100, 200, 300 and 400 MeV/c, respectively. The blue dashed lines 
correspond to the mass spectra for the one-pole SIDD potential 
(SIDD$^{1}$) and the red dash-dotted lines show the mass spectra 
for the two-pole SIDD potential (SIDD$^{2}$). The results 
corresponding to the chiral potential are depicted by black solid 
curves.}
\label{fig.4}
\end{figure*}

The calculated energies of the quasi-bound state are $-24.5$, $-38.3$ and $-43.6$ MeV 
from the $K\bar{K}N$ threshold in the results with chiral, SIDD$^{1}$ and SIDD$^{2}$ 
potentials, respectively. The width of three-hadron decay is estimated to be $45 \sim 110$ MeV. 
Jido {\it et al.}~\cite{d1} made a variational calculation for the three-body $K\bar{K}N$ 
nuclear quasi-bound state using an effective interaction model for $\bar{K}N$, $K\bar{K}$ 
and $KN$ interactions. In this calculation, a quasi-bound state with $I=\frac{1}{2}$ and 
$J^{\pi}=\frac{1}{2}^{+}$, was found with a binding energy about $19 \sim 41$ MeV and a 
width $90 \sim 100$ MeV below the threshold energy of the $K\bar{K}N$ state. The comparison 
of the present results for $K\bar{K}N$ obtained for coupled-channel $\bar{K}N-\pi\Sigma$ and 
$K\bar{K}-\pi\pi-\pi\eta$ interactions with calculations in Ref.~\cite{d1} within the 
variational method and effective $\bar{K}N$ and $K\bar{K}$ interactions shows that they 
are in the same range. However, in the case of the chiral low-energy potential the 
extracted width is smaller than those by other models.

\subsection{Trace of $N^{\star}$ resonance in $(\bar{K}N)_{I=0}+K$ reaction}
The calculated resonance energies presented in Fig.~\ref{fig.3} give only pole 
positions of the three-body $K\bar{K}N$ system. However, we know that 
these results are not a quantity that can be directly measured in any experiment. 
The calculated results in Fig.~\ref{fig.3} could be used as a guideline to study the 
$(\bar{K}N)_{I=0}+K$ reaction. To examine the existence of the quasi-bound state 
in $K\bar{K}N$ system by experiments, one has to calculate the cross sections of 
$K\bar{K}N$ production reactions. As it was said in Sect.~\ref{intro}, the $K\bar{K}N$ 
quasi-bound state can be produced through proton-proton reaction and the trace of 
the resonances would be seen in the mass spectrum of the final particles. In the 
present calculations, I have been studying how well the signature of the $K\bar{K}N$ 
system shows up in the observables of the three-body reactions by using one-channel 
Faddeev equation in the AGS form. To achieve this goal, one must solve the integral 
equations for the amplitudes defined in Eq.~(\ref{eq.4}), and then construct the 
scattering amplitudes defined in Eq.~(\ref{eq.6}). 
\begin{figure*}
\centering
\includegraphics[width=8.cm]{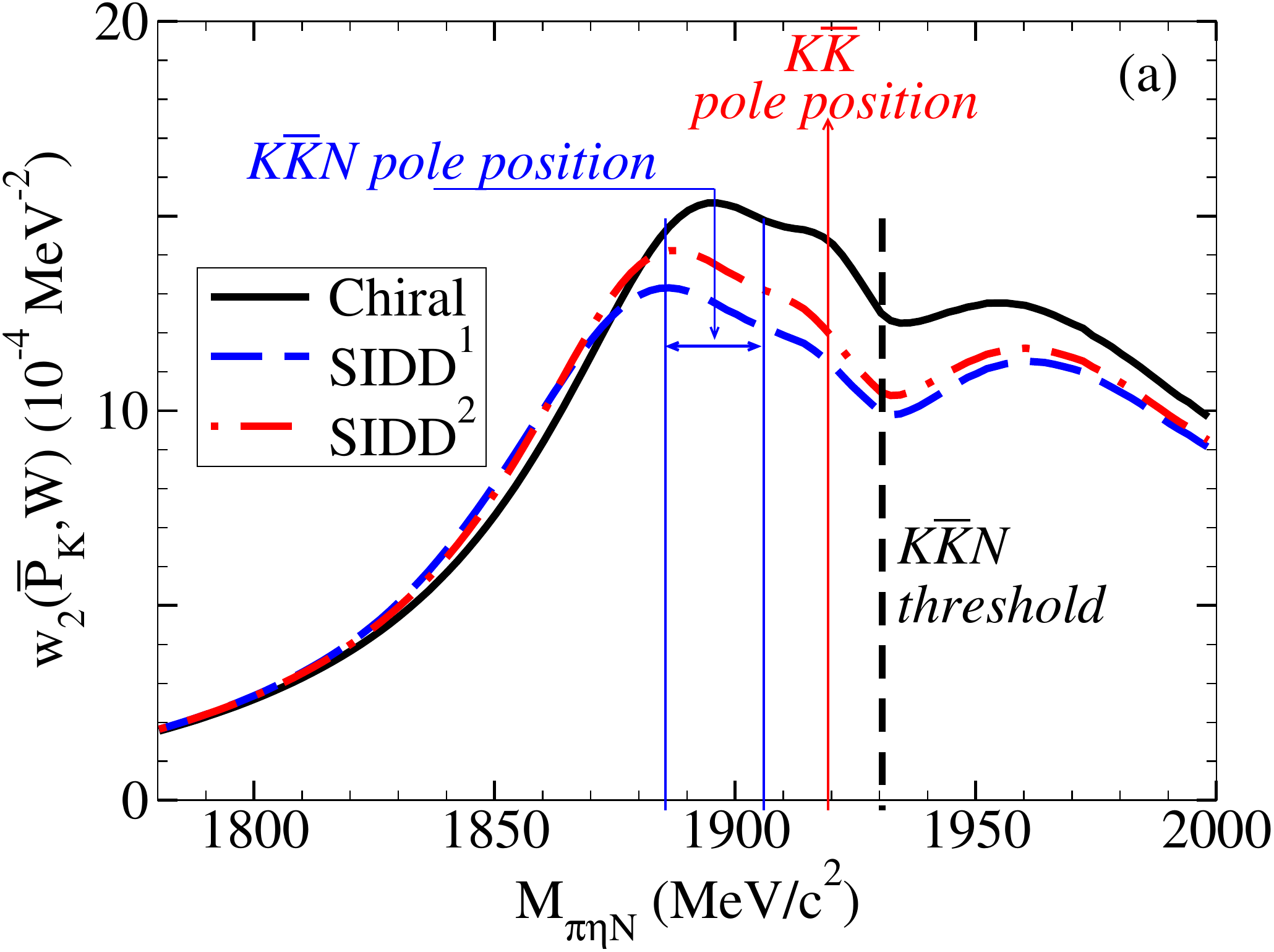}
\includegraphics[width=8.cm]{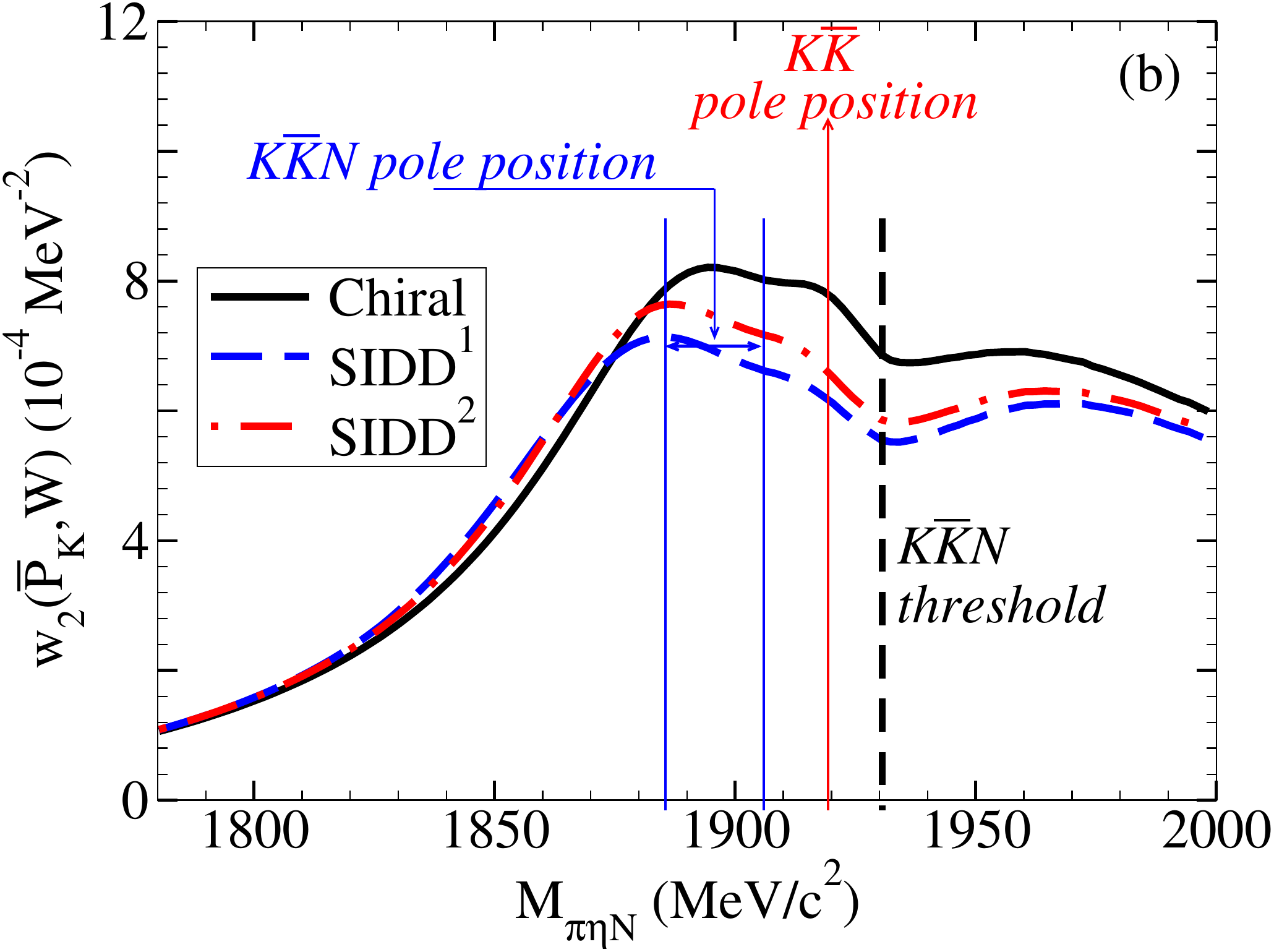} \\ 
\includegraphics[width=8.cm]{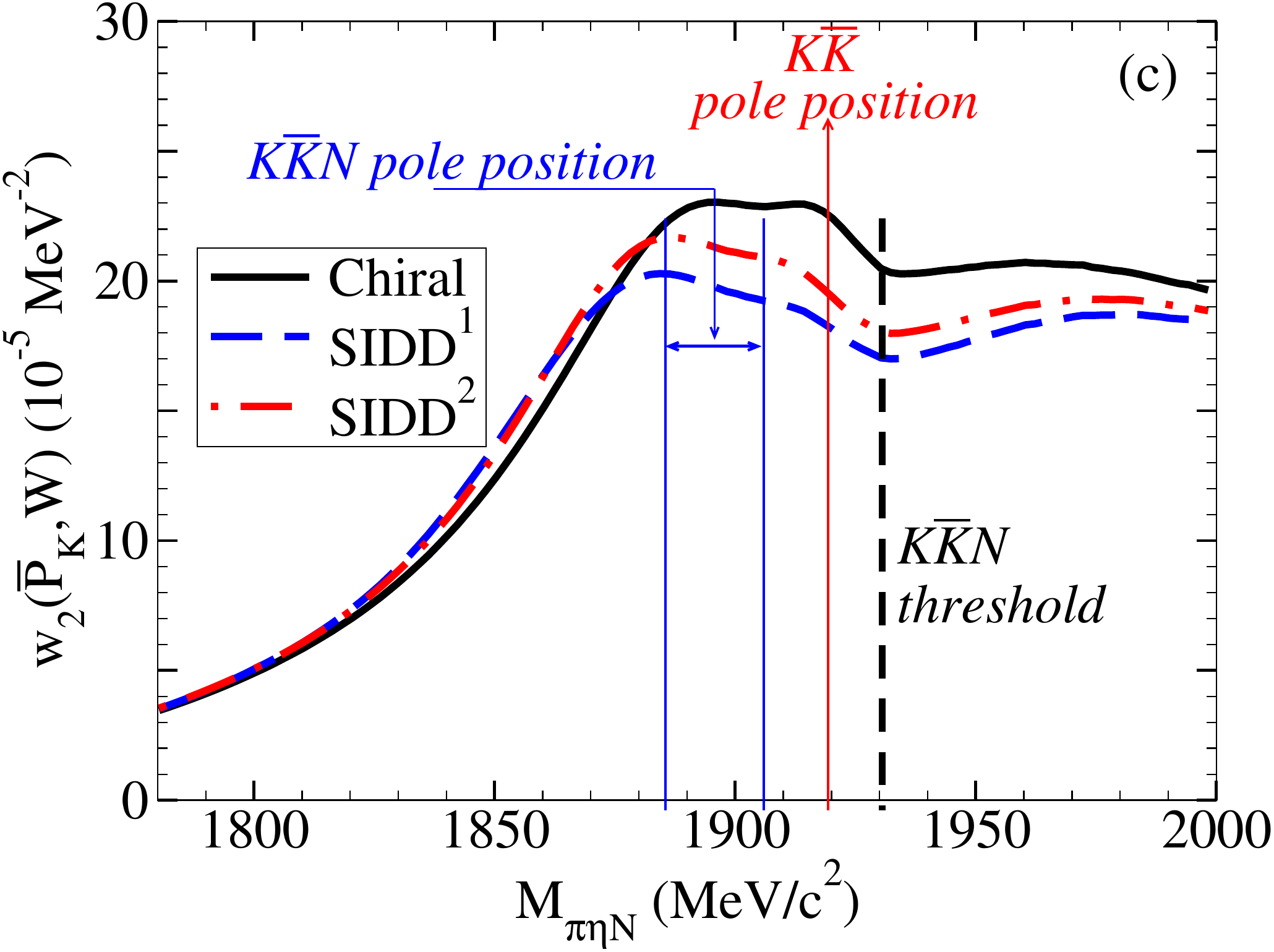}
\includegraphics[width=8.cm]{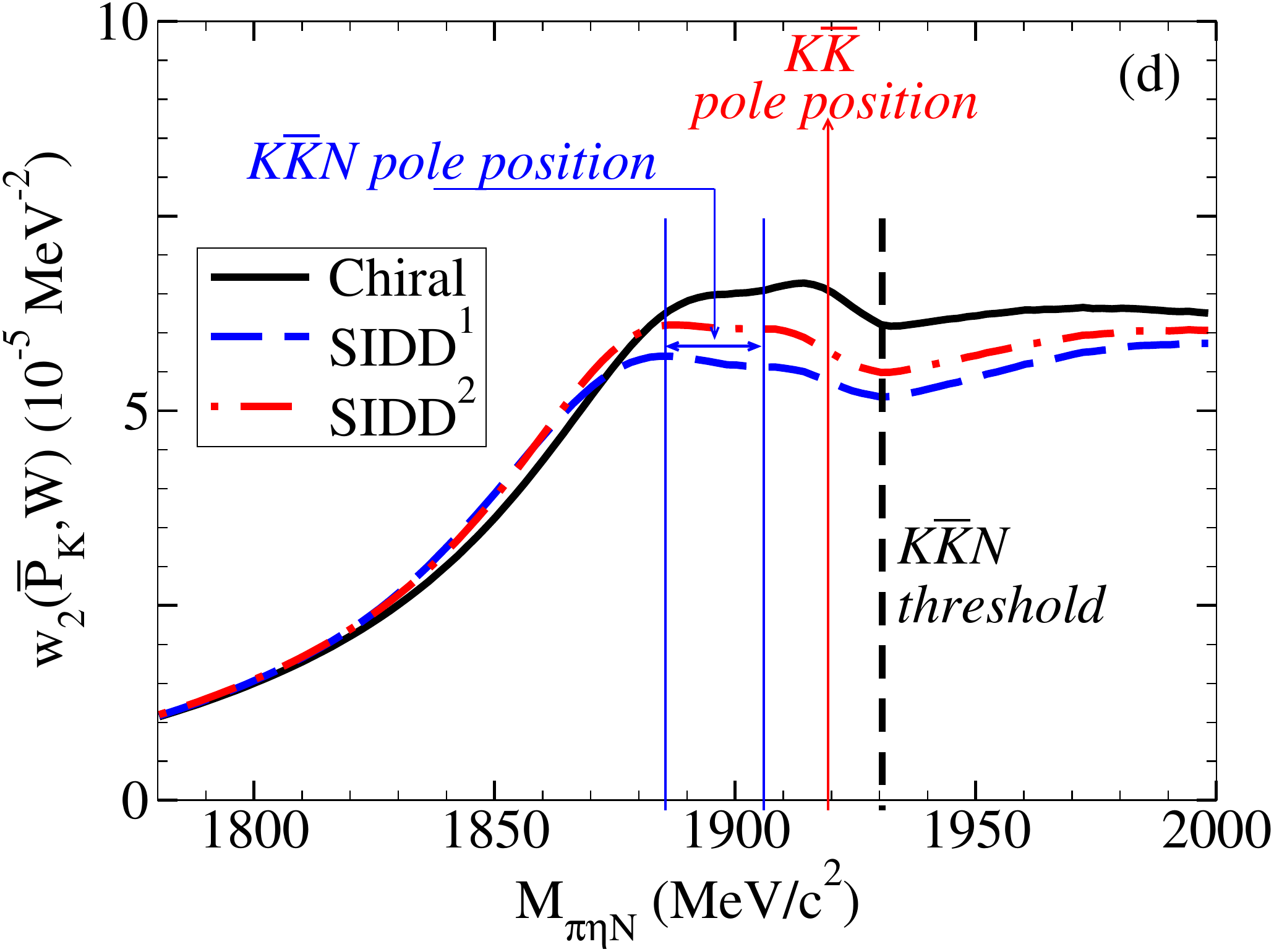}
\caption{(Color online) The $\pi\eta{N}$ mass spectra for 
$(\bar{K}N)_{I=0}+K\rightarrow\pi\eta{N}$ reaction. The 
explanations are same as in Fig.~\ref{fig.4}.}
\label{fig.5}
\end{figure*}

The $\bar{K}N$ is coupled with $\pi\Sigma$ channel and the $K\bar{K}$ system is 
coupled with $\pi\pi$ and $\pi\eta$ channels. Therefore, there are three decay 
channels for $K\bar{K}N$ system, namely, $\pi\Sigma+K$, $\pi\pi+N$ and $\pi\eta+N$. 
In the present study, the first and the third processes are considered for 
$(\bar{K}N)_{I=0}+K$ reaction. To remove the moving singularities in the kernel 
of AGS equations, the so called \textquotedblleft point-method \textquotedblright 
was used. The details of the point-method are given in Refs.~\cite{p1,p2}. 

The transition probabilities for $(\bar{K}N)_{I=0}+K\rightarrow\pi\Sigma{K}$ reaction 
are depicted in Fig.~\ref{fig.4}. To study the dependence of the mass spectra to the 
$\bar{K}N$ interaction, the three-body calculations are performed for chiral based and 
phenomenological potentials for $\bar{K}N$ interaction. To investigate the energy 
dependence of the transition probability, I calculated $w_{1}(\bar{P}_{K},W)$ for 
$\bar{P}_{K}=100-400$ MeV/c. The results suggest that the peak structure in the energy 
region around the $K\bar{K}N$ pole position could be observed, regardless the momentum 
value and the class of the $\bar{K}N-\pi\Sigma$ interaction. In the calculated 
mass spectra for the $\bar{K}N$ potentials having a two-pole structure of $\Lambda(1405)$ 
resonance, the second pole of $\Lambda(1405)$ does not manifest itself the $\pi\Sigma{K}$ 
mass spectra. As one can see from Fig.~\ref{fig.4}, for phenomenological models a second 
peak structure can be seen below the $K\bar{K}N$ threshold. Its position depends on the 
model of interaction and it originates from a branch point in the complex plane~\cite{m1,m2,m3}, 
i.e., a threshold opening associated with the $\Lambda(1405)$ pole.

The results of the Faddeev calculations of the $(\bar{K}N)_{I=0}+K\rightarrow\pi\eta{N}$ 
reaction using different versions of $\bar{K}N-\pi\Sigma$ potentials are shown in 
Fig.~\ref{fig.5}. Within this model, I have found two bump structures appearing in the 
$(\bar{K}N)_{I=0}+K\rightarrow\pi\eta{N}$ transition probabilities in the energy region 
around the $K\bar{K}N$ pole position and $W=M_{K}+M_{\Lambda(1405)}$. Comparing the 
results of the mass spectra with those presented in Fig.~\ref{fig.4}, one can see that 
the peak structures corresponding to the quasi-bound state in $K\bar{K}N$ system is 
more clear and the magnitude of the transition probabilities are 2-4 times bigger than 
those in Fig.~\ref{fig.4}.
\begin{figure*}
\centering
\includegraphics[width=8.cm]{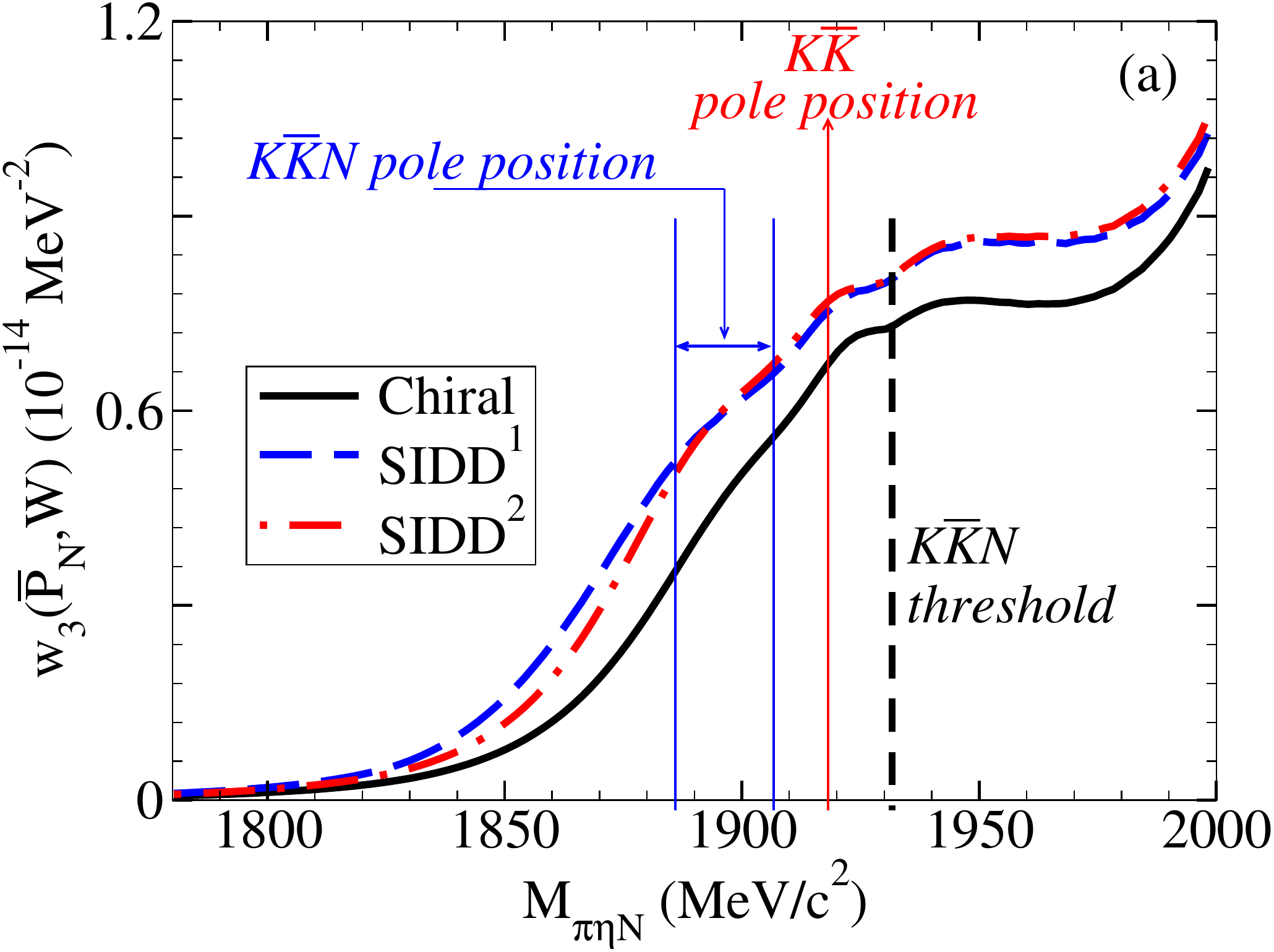}
\includegraphics[width=8.cm]{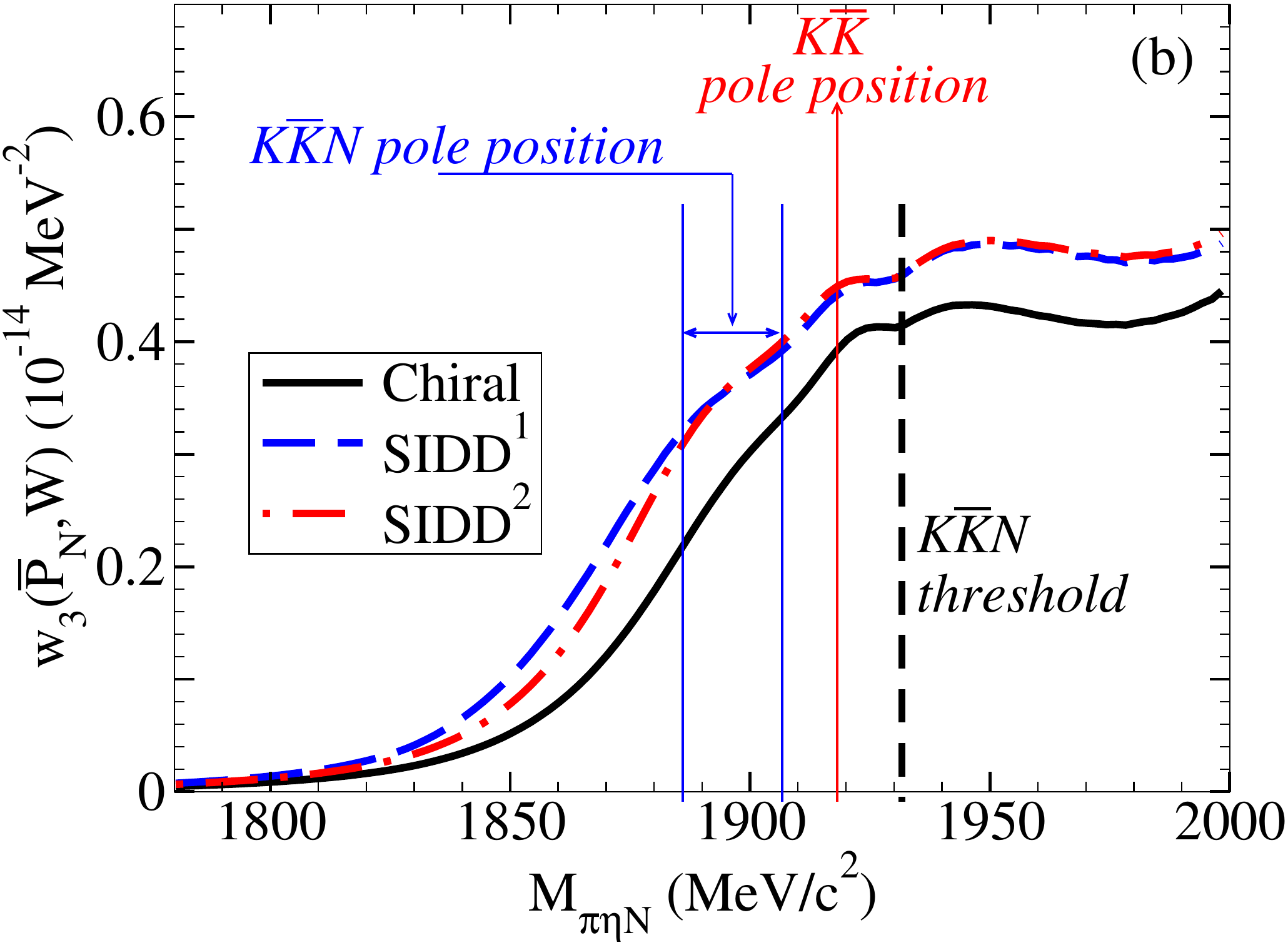} \\
\includegraphics[width=8.cm]{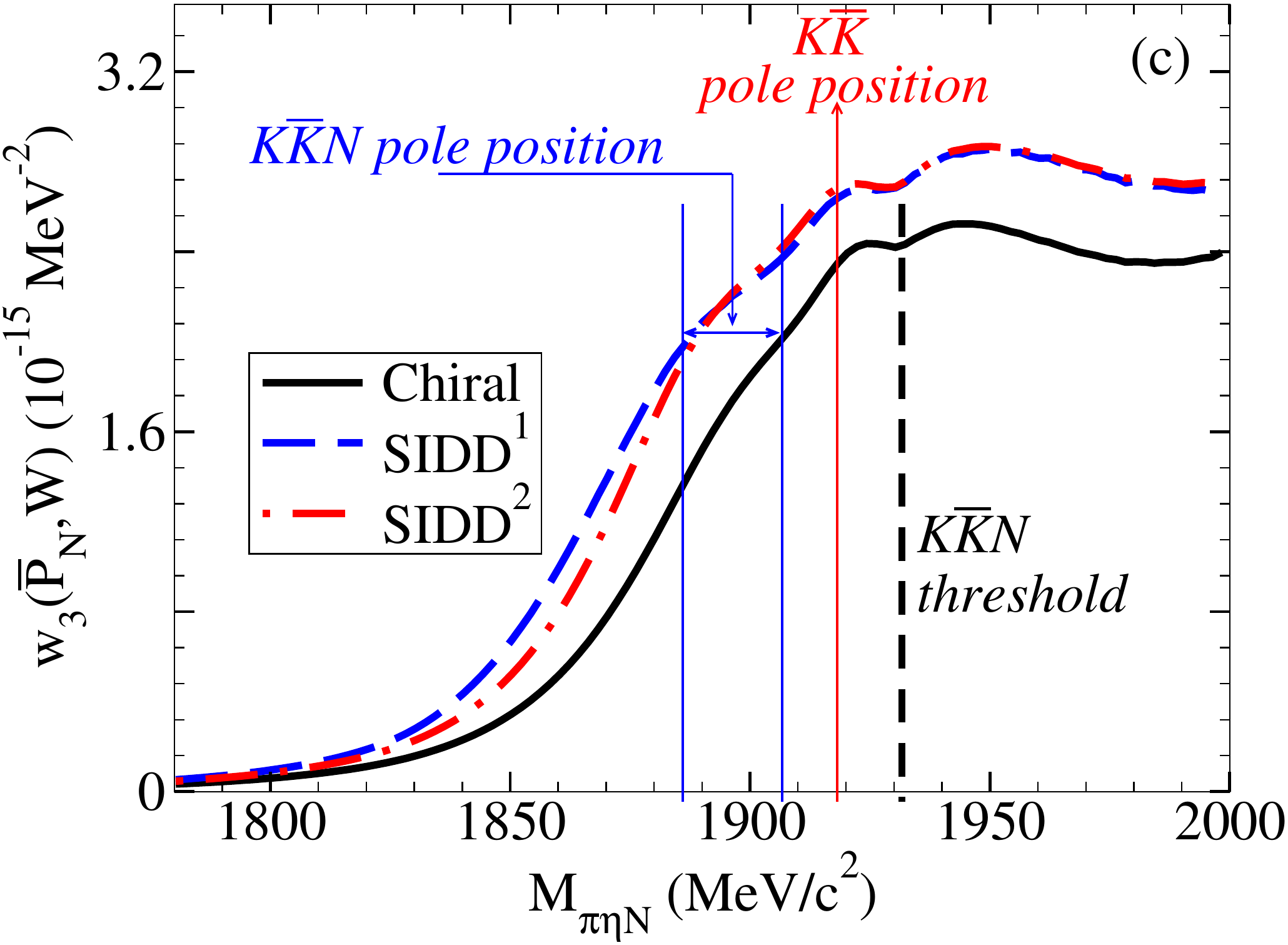}
\includegraphics[width=8.cm]{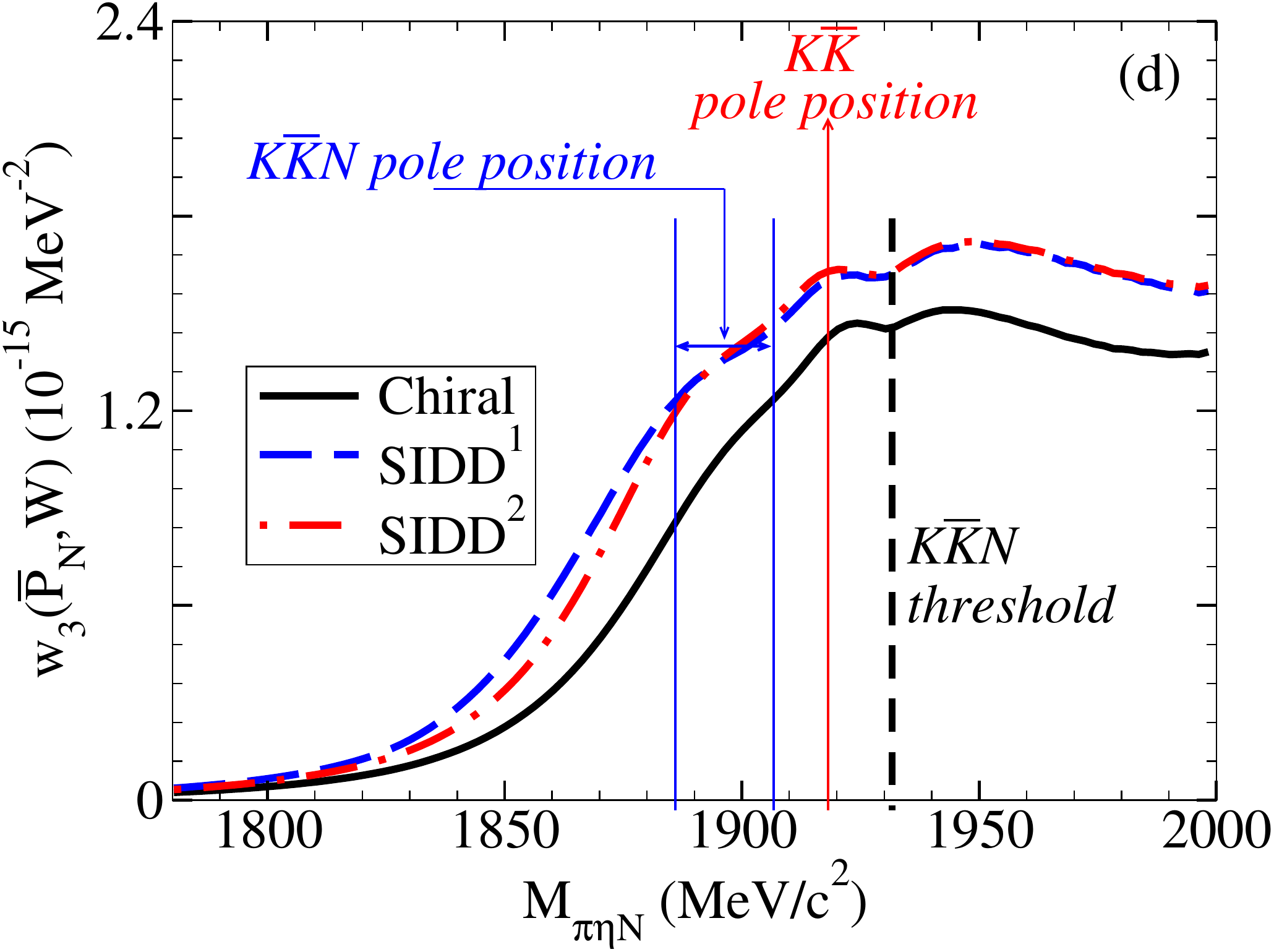}
\caption{(Color online) The $\pi\eta{N}$ mass spectra for 
$(K\bar{K})_{I=1}+N\rightarrow\pi\eta{N}$ reaction. The 
explanations are same as in Fig.~\ref{fig.4}.}
\label{fig.6}
\end{figure*}
\subsection{Trace of $N^{\star}$ resonance in $(K\bar{K})_{I=1}+N$ reaction}
As it was said in Sect.~\ref{intro}, the extracted results in Refs.~\cite{d1,d2,d3} 
suggest that the $K\bar{K}N$ state can be understood by the structure of simultaneous 
coexistence of $\Lambda(1405)+K$ and $a_{0}(980)+N$ clusters and the $\bar{K}$ meson is 
shared by both $\Lambda(1405)$ and $a_{0}(980)$ at the same time. In previous subsection, 
I supposed that the initial structure of the $K\bar{K}N$ system is $(\bar{K}N)_{I=0}+K$ 
and the $\pi\Sigma{K}$ and $\pi\eta{N}$ mass spectra were calculated for $(\bar{K}N)_{I=0}+K$ 
reaction. Now, I suppose that the initial structure of the $K\bar{K}N$ system is 
$(K\bar{K})_{I=1}+N$ and I will study the $\pi\eta{N}$ mass spectra in $(K\bar{K})_{I=1}+N$ 
reaction. Supposing the $(K\bar{K})_{I=1}+N$ as the initial channel of the $K\bar{K}N$ system, 
the three-body Faddeev AGS equations can be given by 
\begin{equation}
\begin{split}
& \mathcal{K}_{K,N}^{I_{{K}},1} = \mathcal{M}_{K,N}^{I_{K},1}  + 
\mathcal{M}_{K,N}^{I_{K},I_{N}}\tau_{N}^{I_{N}}\mathcal{K}_{N,N}^{I_{N},1}
 + \mathcal{M}_{K,\bar{K}}^{I_{K},I_{\bar{K}}}\tau_{\bar{K}}^{I_{\bar{K}}}
\mathcal{K}_{\bar{K},N}^{I_{\bar{K}},1}, \\
& \mathcal{K}_{N,N}^{I_{N},1} =  
\mathcal{M}_{N,K}^{I_{N},I_{K}}\tau_{K}^{I_{K}}
\mathcal{K}_{K,N}^{I_{K},1} + 
\mathcal{M}_{N,\bar{K}}^{I_{N},I_{\bar{K}}}\tau_{\bar{K}}^{I_{\bar{K}}}
\mathcal{K}_{\bar{K},N}^{I_{\bar{K}},1}, \\
& \mathcal{K}_{\bar{K},N}^{I_{\bar{K}},1} =  
\mathcal{M}_{\bar{K},N}^{I_{\bar{K}},1} + 
\mathcal{M}_{\bar{K},K}^{I_{\bar{K}},I_{K}}\tau_{K}^{I_{K}}\mathcal{K}_{K,N}^{I_{K},1} + 
\mathcal{M}_{\bar{K},N}^{I_{\bar{K}},I_{N}}\tau_{N}^{I_{N}}\mathcal{K}_{N,N}^{I_{N},1}, \\
\end{split}
\label{eq.11}
\end{equation}
and consequently, the scattering amplitude for $N+(K\bar{K})_{I=1}\rightarrow\pi\eta+{N}$ 
reaction can be expressed as 
\begin{equation}
\begin{split}
& T_{(\pi\eta)+{N}\leftarrow(K\bar{K})_{I=1}+N} (\vec{k}_{N},\vec{p}_{N},\bar{P}_{N};W)= \\
& \hspace{1.5cm} g_{\pi\eta}^{I=1}(\vec{k}_{N} )\, \tau_{N}^{I=1}(W-\frac{p^{2}_{N}}{2\nu_{N}})
\, \mathcal{K}_{N,N}^{1,1}(p_{N},\bar{P}_{N};W).\\
\end{split}
\label{eq.12}
\end{equation}

Using Eqs.~(\ref{eq.11}) and~(\ref{eq.12}), we define the transition probability of 
$N+(K\bar{K})_{I=1}\rightarrow\pi\eta+{N}$ reaction as follows,
\begin{equation}
\begin{split}
& w_{3}(\bar{P}_{N},W)=\int d^{3}p_{N}\int d^{3}k_{N}\, 
\delta(W-Q_{2}(p_{N},k_{N})) \\
& \hspace{1.7cm}\times 
|T_{(\pi\eta)+{N}\leftarrow(K\bar{K})_{I=1}+N}(\vec{k}_{N},\vec{p}_{N},\bar{P}_{N};W)|^2.
\end{split}
\label{eq.13}
\end{equation}

In Fig.~\ref{fig.6}, the $\pi\eta{N}$ mass spectra for $(K\bar{K})_{I=1}+N\rightarrow\pi\eta{N}$ 
reaction were calculated. The $K\bar{K}N$, $a_{0}(980)+N$ thresholds and also the expected energy 
region for quasi-bound state in $K\bar{K}N$ system are shown using vertical lines. As one can see, 
the mass spectra are affected by two bump structures appearing in the $(K\bar{K})_{I=1}+N\rightarrow\pi\eta{N}$ 
transition probabilities in the energy region around the $K\bar{K}N$ pole position and $z=M_{N}+M_{a_{0}(980)}$, 
where the second bump actually originates from a branch point in the complex plane. Here, the peak 
corresponding to the quasi-bound state in $K\bar{K}N$ is not pronounced in the $\pi\eta{N}$ mass 
spectra as in $(\bar{K}N)_{I=0}+K\rightarrow\pi\eta{N}$ reaction.

In Ref.~\cite{d8}, the possible observation of the $N^{\star}$ was discussed. They provided a series 
of arguments which support the idea that the peak seen in the $\gamma{p}\rightarrow{K}^{+}\Lambda$ 
reaction around 1920 MeV should correspond to the predicted bound state of $K\bar{K}N$ with a mixture 
of $f_{0}(980)$ and $a_{0}(980)$ components. It was said there that an ideal test of the nature of 
the $N^{\star}$ resonance is the study of the $\gamma{p}\rightarrow{K}^{-}K^{+}p$ reaction close to 
threshold. It was concluded that the big asymmetry of the mass distribution with respect to phase 
space close to $M_{inv}=2m_{K}$ is a consequence of the presence of the $f_{0}(980)$ or $a_{0}(980)$ 
below threshold. Furthermore, it was observed that the $\gamma{p}\rightarrow{K}^{-}K^{+}p $ cross 
section is more pronounced at lower energies which is a consequence of the presence of the 
three-particle resonance below threshold. In present paper, it was also shown that one can see the 
signal of the $K\bar{K}N$ quasi-bound state in mass spectrum of the final particles and also the 
mass spectra are affected by branch points resulting from the resonances in two-body subsystems. 
\section{CONCLUSION}
\label{conc}
In summary, the homogeneous Faddeev AGS equation for $K\bar{K}N$ system was solved and the pole 
position of $K\bar{K}N$ system for different types of $\bar{K}N-\pi\Sigma$ potentials was calculated. 
The transition probabilities for $(\bar{K}N)_{I=0}+K$ reaction were extracted and the possible 
observation of $N^{\star}$ in mass spectrum of the decay products was studied. Based on Faddeev 
approach, the $\pi\Sigma{K}$ and $\pi\eta{N}$ mass spectra for different types of $\bar{K}N$ 
and $K\bar{K}$ interactions were calculated. Within this model, it was found a bump produced 
by $K\bar{K}N$ system appearing in the $(\bar{K}N)_{I=0}+K$ transition probabilities in energy 
region around the $K\bar{K}N$ pole position for momentum $\bar{P}_{K}=100-400$ MeV/c. Furthermore, 
it was observed that the shape and position of the peaks in the transition probabilities are 
independent of the momentum $\bar{P}_{K}$ of the initial $(\bar{K}N)_{I=0}+K$ channel. 
It was shown that not only one can see the signature of the $K\bar{K}N$ quasi-bound state, 
but also, one can see the effect of the branch points which are resulting from $\Lambda$(1405), 
$f_{0}(980)$ and $a_{0}(980)$ poles. The bump structures related to the branch points can 
affect the peak corresponding to the $K\bar{K}N$ quasi-bound state. Therefore, this reaction 
would also be helpful to reveal the dynamical origin of two-body resonances. 
The $K\bar{K}N$ system is mainly dominated by $\Lambda(1405)+K$ and $a_{0}(980)+N$ structures. 
Therefore, the $(K\bar{K})_{I=1}+N\rightarrow\pi\eta{N}$ reaction was also investigated and it 
was observed that the $\pi\eta{N}$ mass spectrum reveals the same behavior as in the case of 
$(\bar{K}N)_{I=0}+K$ reaction. However, the magnitude of the extracted mass spectra are 
considerably smaller than those resulting from $(\bar{K}N)_{I=0}+K$ reaction.

\end{document}